\documentclass{article}
\usepackage[T1]{fontenc} % add special characters (e.g., umlaute)
\usepackage[utf8]{inputenc} % set utf-8 as default input encoding
\usepackage{ismir,amsmath,cite,url}
\usepackage{graphicx}
\usepackage{color}
\usepackage{enumitem}
\usepackage{comment}

\usepackage{lineno}
\usepackage{multirow}
\usepackage{amssymb}
\title{Automatic Piano Transcription with Hierarchical Frequency-Time Transformer}

\multauthor
{Keisuke Toyama$^1$ \hspace{1cm} Taketo Akama$^2$ \hspace{1cm} Yukara Ikemiya$^1$} { \bfseries{Yuhta Takida$^1$ \hspace{1cm} Wei-Hsiang Liao$^1$ \hspace{1cm} Yuki Mitsufuji$^1$}\\
  $^1$ Sony Group Corporation, Japan\\
$^2$ Sony Computer Science Laboratories, Japan\\
{\tt\small keisuke.toyama@sony.com}
}
\def\authorname{K. Toyama, T. Akama, Y. Ikemiya, Y. Takida, W. H. Liao, and Y. Mitsufuji}
\begin{document}

\maketitle
%
% Abstract
%
\begin{abstract}
Taking long-term spectral and temporal dependencies into account is essential for automatic piano transcription.
This is especially helpful when determining the precise onset and offset for each note in the polyphonic piano content.
In this case, we may rely on the capability of self-attention mechanism in Transformers to capture these long-term dependencies in the frequency and time axes.
In this work, we propose \textit{hFT-Transformer}, which is an automatic music transcription method that uses a two-level hierarchical frequency-time Transformer architecture.
The first hierarchy includes a convolutional block in the time axis, a Transformer encoder in the frequency axis, and a Transformer decoder that converts the dimension in the frequency axis.
The output is then fed into the second hierarchy which consists of another Transformer encoder in the time axis.
We evaluated our method with the widely used MAPS and MAESTRO v3.0.0 datasets, and it demonstrated state-of-the-art performance on all the F1-scores of the metrics among \textit{Frame}, \textit{Note}, \textit{Note with Offset}, and \textit{Note with Offset and Velocity} estimations.
\end{abstract}

%
% 1. Introduction
%
\section{Introduction}\label{sec:introduction}
\begin{figure*}
 \centering
 \includegraphics[width=0.99\linewidth]{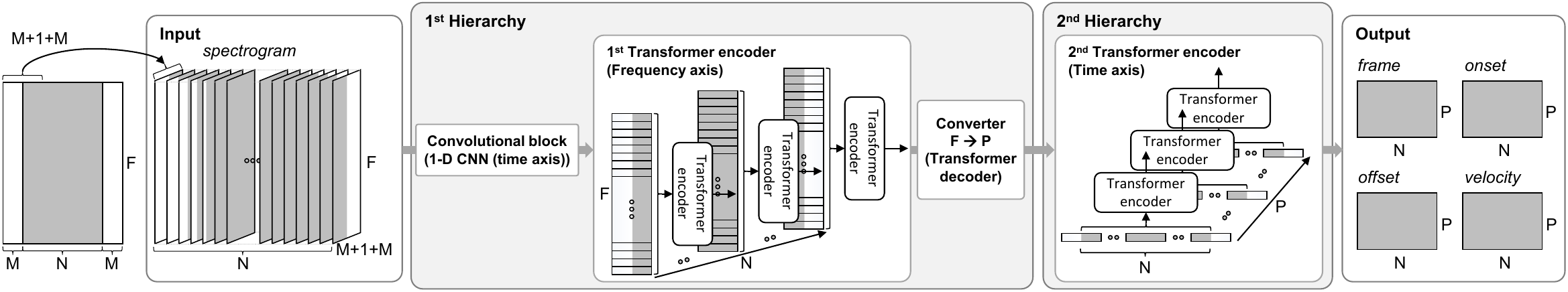}
 \caption{hFT-Transformer (N: number of frames in each processing chunk, M: length of margin, F: number of frequency bins, P: number of pitches)}
 \label{fig:figure_1}%Figure 1
\end{figure*}
 
Automatic music transcription (AMT) is to convert music signals into symbolic representations such as piano rolls, Musical Instrument Digital Interface (MIDI), and musical scores~\cite{BENETOS2019:01}.
AMT is important for music information retrieval (MIR), its result is useful for symbolic music composition, chord progression recognition, score alignment, etc.
Following the conventional methods~\cite{BENETOS2019:01,OU2022:01,HAWTHORNE2021:01,SIGTIA2016:01,WEI2022:01,KONG2021:01,HAWTHORNE2018:01,KWON2020:01,CHEUK2020:01,KELZ2019:01,GARDNER2022:01,YAN2021:01,HAWTHORNE2019:01,KELZ2016:01,KIM2019:01}, we estimate the frame-level metric and note-level metrics as follows: (1) \textit{Frame}: the activation of quantized pitches in each time-processing frame, (2) \textit{Note}: the onset time of each note, (3) \textit{Note with Offset}: the onset and offset time of each note, and (4) \textit{Note with Offset and Velocity}: the onset, offset time, and the loudness of each note.

For automatic piano transcription, it is important to analyze several harmonic structures that spread in a wide range of frequencies, since piano excerpts are usually polyphonic.
Convolutional neural network (CNN)-based methods have been used to aggregate harmonic structures as acoustic features.
Most conventional methods apply multi-layer convolutional blocks to extend the receptive field in the frequency axis.
However, the blocks often include pooling or striding to downsample the features in the frequency axis.
Such a downsampling process may reduce the frequency resolution~\cite{KONG2021:01}.
It is worth mentioning, many of these methods use 2-D convolutions, which means the convolution is simultaneously applied in the frequency and time axes.
The convolution 
in the time axis works as a pre-emphasis filter to model the temporal changes of the input signals.

Up to now, recurrent neural networks (RNNs), such as gated recurrent unit (GRU)~\cite{KYUNGHYUN2014:01} and long short-term memory (LSTM)~\cite{HOCHREITER1997:01}, are popular for analyzing the temporal sequences of acoustic features.
However, recently some of the works start to use Transformer~\cite{VASWANI2017:01}, which is a powerful tool for analyzing sequences, in AMT tasks.
Ou et al.~\cite{OU2022:01} applied a Transformer encoder along the time axis and suggested that using Transformer improves velocity estimation.
Hawthorne et al.~\cite{HAWTHORNE2021:01} used a Transformer encoder-decoder as a sequence-to-sequence model for estimating a sequence of note events from another sequence of input audio spectrograms.
Their method outperformed other methods using GRUs or LSTMs.
Lu et al.~\cite{LU2021:01} proposed a method called SpecTNT to apply Transformer encoders in both frequency and time axes and reached state-of-the-art performance for various MIR tasks such as music tagging, vocal melody extraction, and chord recognition.
This suggests that such a combination of encoders helps in characterizing the broad-scale dependency in the frequency and time axes.
However, SpecTNT aggregates spectral features into one token, and the process in its temporal Transformer encoder is not independent in the frequency axis. 
This inspires us to incorporate Transformer encoders in the frequency and time axes and make the spectral information available for the temporal Transformer encoder.

In addition, we usually divide the input signal into chunks since the entire sequence is often too long to be dealt at once.
However, this raises a problem that the estimated onset and offset accuracy fluctuates depending on the relative position in the processing chunk.
In our observation, the accuracy tends to be worse at both ends of the processing chunk.
This motivates us to incorporate extra techniques during the inference time to boost the performance. 

In summary, we propose \textit{hFT-Transformer}, an automatic piano transcription method that uses a two-level hierarchical frequency-time Transformer architecture.
Its workflow is shown in Figure \ref{fig:figure_1}.
The first hierarchy consists of a one-dimensional (1-D) convolutional block in the time axis, a Transformer encoder in the frequency axis, and a Transformer decoder in the frequency axis.
The second hierarchy consists of another Transformer encoder in the time axis.
In particular, the Transformer decoder at the end of the first hierarchy converts the dimension in the frequency axis from the number of frequency bins to the number of pitches (88 for piano).
Regarding the issue of the location dependent accuracy fluctuation in the processing chunks, we propose a technique which halves the stride length at inference time.
It uses only the result of the central part of processing chunks, which will improve overall accuracy.
Finally, in Section \ref{sec:experiments}, we show that our method outperforms other piano transcription methods in terms of F1 scores for all the four metrics.

A \texttt{PyTorch} implementation of our method is available here\footnote{\texttt{https://github.com/sony/hFT-Transformer}}.

%
% 2. Related Works
%
\section{Related Work}\label{sec:related_work}
Neural networks, such as CNNs, RNNs, generative adversarial networks (GANs)~\cite{GOODFELLOW2014:01}, and Transformers have been dominant for AMT.
Since Sigtia et al.~\cite{SIGTIA2016:01} proposed the first method to use a CNN to tackle AMT, CNNs have been widely used for the methods of analyzing the spectral dependency of the input spectrogram~\cite{KELZ2016:01,HAWTHORNE2018:01,HAWTHORNE2019:01,KIM2019:01,KELZ2019:01,KWON2020:01,CHEUK2020:01,KONG2021:01,YAN2021:01,OU2022:01}.
However, it is difficult for CNNs to directly capture the harmonic structure of the input sound in a wide range of frequencies, as convolutions are used to capture features in a local area.
Wei et al.~\cite{WEI2022:01} proposed a method of using harmonic constant-Q transform (CQT) for capturing the harmonic structure of piano sounds.
They first applied a 3-Dimensional CQT,
then applied multiple dilated convolutions with different dilation rates to the output of CQT.
Because the dilation rates are designed to capture the harmonics, the performance of \textit{Frame} and \textit{Note} accuracy reached state-of-the-art.
However, the dilation rates are designed specifically for piano.
Thus, the method is not easy to adapt to other instruments.

For analysis of time dependency, Kong et al.~\cite{KONG2021:01} proposed a method that uses GRUs.
Howthorner et al.~\cite{HAWTHORNE2018:01}, Kwon et al.~\cite{KWON2020:01}, Cheuk et al.~\cite{CHEUK2020:01}, and Wei et al.~\cite{WEI2022:01} proposed methods that use bi-directional LSTMs for analysis.
Ou et al.~\cite{OU2022:01} used a Transformer encoder to replace the GRUs in Kong et al.'s method~\cite{KONG2021:01}, and showed the effectiveness of the Transformer.
Usually, the note onset and offset are estimated in each frequency and time-processing frame grid, then paired as a note for note-level transcription by post-processing algorithms such as \cite{KONG2021:01}.
However, compared to heuristically designed algorithms, end-to-end data-driven methods are often preferred.
For example, Keltz et al.~\cite{KELZ2019:01} applied a seven-state hidden Markov model (HMM) for the sequence of attack, decay, sustain, and release to achieve note-level transcription.
Kwon et al.~\cite{KWON2020:01} proposed a method of characterizing the output of LSTM as a five-state statement (onset, offset, re-onset, activate, and inactivate).
Hawthorne et al.~\cite{HAWTHORNE2021:01} proposed a method of estimating a sequence of note events, such as note pitch, velocity, and time, from another sequence of input audio spectrograms using a Transformer encoder-decoder. 
This method performs well in multiple instruments with the same model~\cite{GARDNER2022:01}.
Yan et al.~\cite{YAN2021:01} proposed a note-wise transcription method for estimating the interval between onset and offset.
This method shows state-of-the-art performance in estimating \textit{Note with Offset} and \textit{Note with Offset and Velocity}.
However, the performance in estimating \textit{Frame} and \textit{Note} is worse than that of Wei et al.'s method~\cite{WEI2022:01}.

\begin{figure*}[t]
 \centering
 \includegraphics[width=0.82\linewidth]{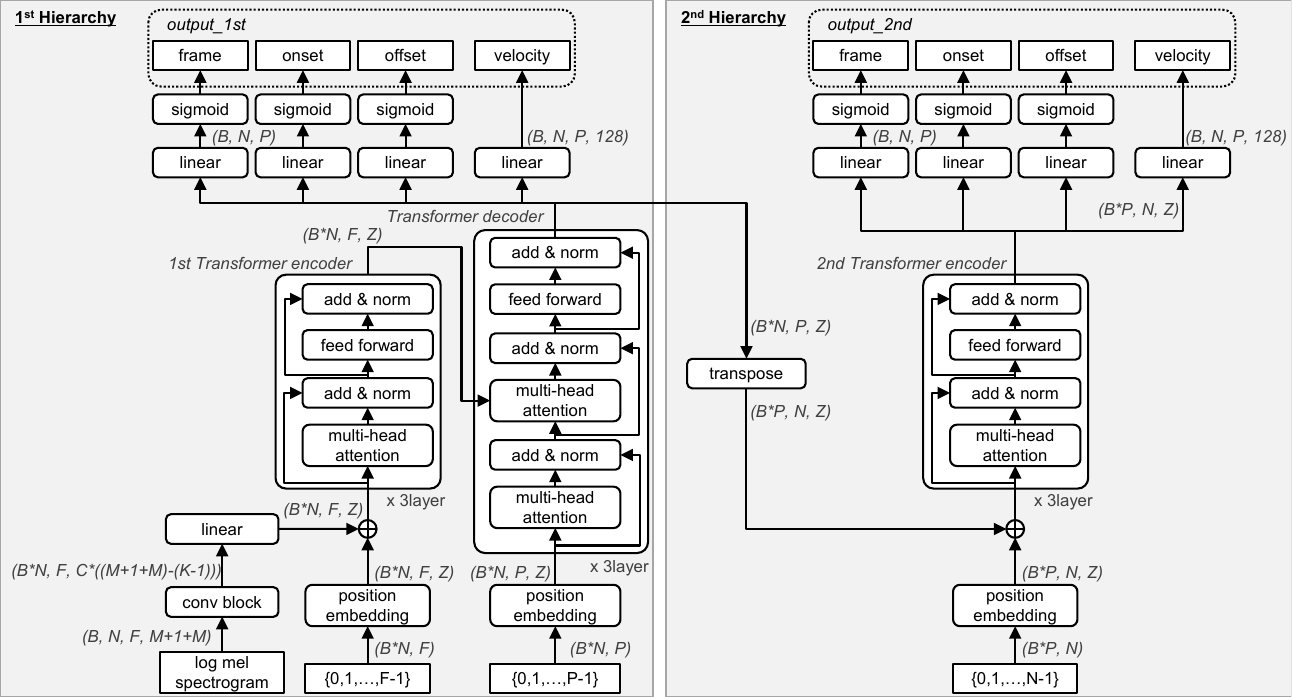}
 \caption{Model architecture of hFT-Transformer}
 \label{fig:figure_2}%Figure 2
\end{figure*}

%
% 3. Method
%
\section{Method}\label{sec:method}

%
% 3-1. Configuration
%
\subsection{Configuration}\label{subsec:configuration}
Our proposed method aims to transcribe $N$ frames of the input spectrogram into $N$ frames of the output piano rolls (\textit{frame}, \textit{onset}, \textit{offset}, and \textit{velocity}) as shown in Figure \ref{fig:figure_1}, where $N$ is the number of frames in each processing chunk.
Each input frame is composed of a log-mel spectrogram having size ($F$, $M+1+M$), where $F$ is the number of frequency bins, and $M$ is the size of the forward margin and that of the backward margin.
To obtain the log-mel spectrogram, we first downmix the input waveform into one channel and resample them to 16 kHz.
Then, the resampled waveform is transformed into a mel spectrogram with \texttt{transforms.MelSpectrogram} class in the \texttt{Torchaudio} library~\cite{yang2021torchaudio}.
For the transformation, we use \textit{hann} window, setting the window size as 2048, fast-Fourier-transform size as 2048, $F$ as 256, padding mode as \textit{constant}, and hop-size as 16 ms.
The magnitude of the mel spectrogram is then compressed with a log function.

%
% 3-2. Model Architectue and Loss Functions
%
\subsection{Model Architecture and Loss Functions}\label{subsec:model_architecture_and_loss_functions}
The model architecture of our proposed method is shown in Figure \ref{fig:figure_2}.
We first apply a convolutional block to the input log-mel spectrogram, the size of which is ($B$, $N$, $F$, $M+1+M$) where $B$ is the batch size.
In the convolutional block, we apply a 1-D convolution in the $M+1+M$ dimension.
After this process, the data are embedded with a linear module.

The embedded vector is then processed with the first Transformer encoder in the frequency axis.
The self-attention is processed to analyze the dependency between spectral features.
The positional information is designated as [$0$, $1$, ..., $F-1$].
These positional values are then embedded with a trainable embedding.
These are processed in the frequency axis only, thus completely independent to the time axis ($N$ dimension).

Next, we convert the frequency dimension from $F$ to the number of pitches ($P$).
A Transformer decoder with cross-attention is used as the converter.
The Transformer decoder calculates the cross-attention between the output vectors of the first Transformer encoder and another trainable positional embedding made from [$0$, $1$, ..., $P-1$].
%The positional values are embedded with a trainable embedding.
The decoded vectors are then converted to the outputs of the first hierarchy with a linear module and a sigmoid function (hereafter, we call these outputs \textit{output\_1st}).

Regarding the loss calculation for the outputs, \textit{frame}, \textit{onset}, and \textit{offset} are calculated with binary cross-entropy, and \textit{velocity} is calculated with 128-category cross-entropy.
The losses can be summarized as the following equations:
\begin{linenomath}
\begin{align}
    L_{\mathrm{bce}}^{\mathrm{<m>}}&=\sum_{n=0}^{N-1}\sum_{p=0}^{P-1}l_{\mathrm{bce}}(y_{n,p}^{\mathrm{<m>}},\hat{y}_{n,p}^{\mathrm{<m>}}),\\
    L_{\mathrm{cce}}^{\mathrm{velocity}}&=\sum_{n=0}^{N-1}\sum_{p=0}^{P-1}l_{\mathrm{cce}}(y_{n,p}^{\mathrm{velocity}},\hat{y}_{n,p}^{\mathrm{velocity}}),\\
    L&=L_{\mathrm{bce}}^{\mathrm{frame}}+L_{\mathrm{bce}}^{\mathrm{onset}}+L_{\mathrm{bce}}^{\mathrm{offset}}+L_{\mathrm{cce}}^{\mathrm{velocity}}\label{eqn:loss_1st},
\end{align}
\end{linenomath}
where $\mathrm{<m>}$ is the placeholder for each output (\textit{frame}, \textit{onset}, and \textit{offset}), $l_{\mathrm{bce}}$ and $l_{\mathrm{cce}}$ denote the loss function for binary cross-entropy and categorical cross-entropy, respectively, and $y$ and $\hat{y}$ denote the ground truth and predicted values of each output (\textit{frame}, \textit{onset}, \textit{offset}, and \textit{velocity}), respectively.
Although it is intuitive to apply the mean squared error (MSE) for \textit{velocity}, we found that using the categorical cross-entropy yields much better performance than the MSE from a preliminary experiment.

Finally, the output of the converter is processed with another Transformer encoder in the time axis.
The self-attention is used to analyze the temporal dependency of features in each time-processing frame.
A third positional embedding made from [$0$, $1$, ..., $N-1$] is used here.
Then, similar to the first hierarchy, the outputs of the second hierarchy are obtained through a linear module and a sigmoid function.
We call these outputs of the second hierarchy as \textit{output\_2nd} hereafter.
The losses for the \textit{output\_2nd} are evaluated in the same way as those for \textit{output\_1st}.
These losses are summed with the coefficients $\alpha_{\mathrm{1st}}$ and $\alpha_{\mathrm{2nd}}$ as follows:
\begin{linenomath}
\begin{align}
    L_{\mathrm{all}}=\alpha_{\mathrm{1st}}L_{\mathrm{1st}}+\alpha_{\mathrm{2nd}}L_{\mathrm{2nd}}.\label{eqn:loss_combination}
\end{align}
\end{linenomath}
Although both outputs are used for computing losses during training, only \textit{output\_2nd} is used in inference.
As Chen et al.~\cite{CHEN-Yu-Hua2022:01} suggested that the performance of their method of calculating multiple losses outperformed the method that uses single loss only, it hints us that utilizing both \textit{output\_1st} and \textit{output\_2nd} in training has the potential to achieve better performance.

%
% 3-3. Stride for Inference
%
\subsection{Inference Stride}\label{subsec:inference_stride}
\begin{figure}[t]
 \centering
 \includegraphics[width=0.99\columnwidth]{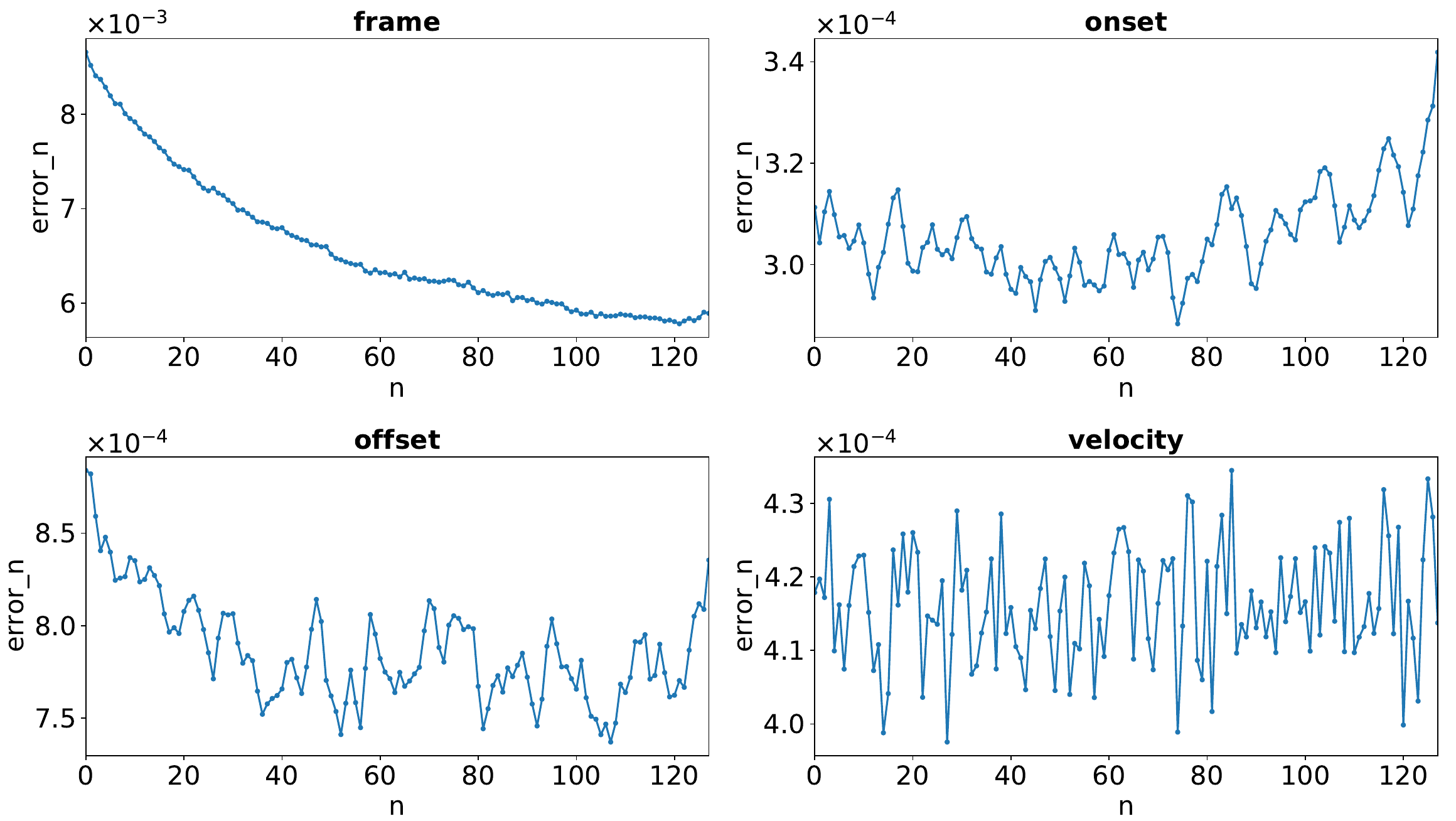}
 \caption{Estimation error (Eqn (\ref{eqn:eight})) on location in each time-processing frame}
 \label{fig:figure_3}%Figure 3
\end{figure}

\begin{figure}[t]
 \centering
 \includegraphics[width=0.7\columnwidth]{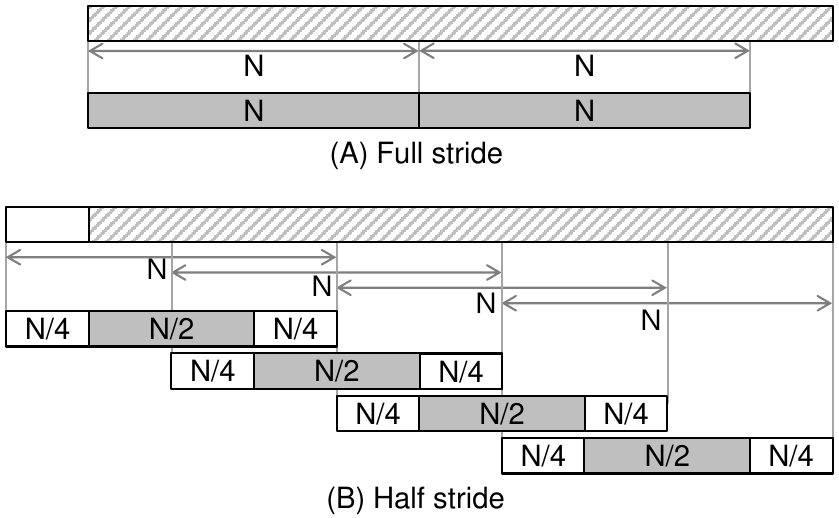}
 \caption{Inference stride: (A) full stride, (B) half stride}
 \label{fig:figure_4}%Figure 4
\end{figure}

As mentioned in Section \ref{sec:introduction}, chunk-based processing is required because the input length is limited due to system limitations, such as memory size and acceptable processing delay. 
We found that the estimation error tends to increase at certain part within each processing chunk.
This can be demonstrated by evaluating the error for each instance of time $n$ within the chunks:
\begin{equation}
    \mathit{error}_{n}^{\mathrm{<m>}}=\frac{1}{IP}\sum_{i=0}^{I-1}\sum_{p=0}^{P-1}(y_{i,n,p}^{\mathrm{<m>}}-\hat{y}_{i,n,p}^{\mathrm{<m>}})^2,
    \label{eqn:eight}
\end{equation}
where $\mathrm{<m>}$ is the placeholder for each output (\textit{frame}, \textit{onset}, \textit{offset}, and \textit{velocity}), and $I$ is the number of processing chunks over the test set.
The result using our proposed model trained using the MAESTRO training set (described in Section \ref{sec:experiments}) is shown in Figure \ref{fig:figure_3}.
Here, the error $\mathit{error}_{n}^{\mathrm{<m>}}$ is calculated using the MAESTRO test set.
In the figure, we observe a monotonic decrease for \textit{frame} and a similar but much weaker trend for \textit{onset} and \textit{offset}. However, for \textit{velocity}, no such trend can be observed.
This hints us to use only the middle portion of a processing chunk as the output to reduce the error rate. We call this as the half-stride strategy, since a $50\%$ overlap is required for processing chunks, as shown in Figure \ref{fig:figure_4} (B).

%
% 4. Experiments
%
\section{Experiments}\label{sec:experiments}
%
% 4-1. Dataset
%
\subsection{Datasets}\label{subsec:datasets}
We use two well-known piano datasets for the evaluation.
The MAPS dataset~\cite{EMIYA2009:01} consists of CD-quality recordings and corresponding annotations of isolated notes, chords, and complete piano pieces.
We use the full musical pieces and the train/validation/test split as stated in \cite{SIGTIA2016:01, HAWTHORNE2018:01}.
The number of recordings and the total duration in hours in each split are 139/71/60 and 8.3/4.4/5.5, respectively.
The MAESTRO v3.0.0 dataset~\cite{HAWTHORNE2019:01} includes about 200 hours of paired audio and MIDI recordings from ten years of the International Piano-e-Competition.
We used the train/validation/test split configuration as provided.
In each split, the number of recordings and total duration in hours are 962/137/177 and 159.2/19.4/20.0, respectively.
For both datasets, the MIDI data have been collected by Yamaha Disklaviers concert-quality acoustic grand pianos integrated with a high-precision MIDI capture and playback system.

%
% 4-2. Model Configuration
%
\subsection{Model Configuration}\label{subsec:model_configuration}
Regarding our model architecture depicted in Figure \ref{fig:figure_2}, we set $N$ as 128, $M$ as 32, $F$ as 256, $P$ as 88, the CNN channels ($C$) as 4, size of the CNN kernel ($K$) as 5, and embedding vector size ($Z$) as 256.
For the Transformers, we set the feed-forward network vector size as 512, the number of heads as 4, and the number of layers as 3.
For training, we used the following settings: a batch size of 8, learning rate of $0.0001$ with Adam optimizer\cite{KINGMA2015:01}, dropout rate of $0.1$, and clip norm of $1.0$.
\texttt{ReduceLROnPlateu} in \texttt{PyTorch} is used for learning rate scheduling with default parameters.
We set $\alpha_{\mathrm{1st}}$ and $\alpha_{\mathrm{2nd}}$ as 1.0, which were derived from a preliminary experiment (see Section \ref{subsec:ablation_study}).

We trained our models for 50 epochs on MAPS dataset and 20 epochs for MAESTRO dataset using one NVIDIA A100 GPU.
It took roughly 140 minutes and 43.5 hours to train one epoch with our model for MAPS and MAESTRO, respectively.
The best model is determined by choosing the one with the highest F1 score in the validation stage.

In order to obtain high-resolution ground truth for \textit{onset} and \textit{offset}, we followed the method in Kong et al.~\cite{KONG2021:01}.
We set $J$, the hyper-parameter to control the sharpness of the targets, to 3.
Also, the label of \textit{velocity} is set only when an \textit{onset} is present.
We set the threshold as $0.5$, which means if the \textit{onset} is smaller than $0.5$, the \textit{velocity} is set as 0.

%
% 4-3. Inference
%
\begin{table*}[t]
 \begin{center}
 \small
 \tabcolsep=3pt
 \begin{tabular}{ccc|cccccccccccc}
  \hline
  \multirow{2}{*}{Method}&half&\multirow{2}{*}{Params}&\multicolumn{3}{c}{Frame}&\multicolumn{3}{c}{Note}&\multicolumn{3}{c}{Note w/ Offset}&\multicolumn{3}{c}{Note w/ Offset\&Velocity}\\
  &stride&&P(\%)&R(\%)&F1(\%)&P(\%)&R(\%)&F1(\%)&P(\%)&R(\%)&F1(\%)&P(\%)&R(\%)&F1(\%)\\
  \hline
  Onsets\&Frames\cite{HAWTHORNE2018:01}&&26M&\underline{88.53}&70.89&78.30&84.24&80.67&82.29&51.32&49.31&50.22&35.52&30.80&35.59\\
  ADSR\cite{KELZ2019:01}&&0.3M&\textbf{90.73}&67.85&77.16&\textbf{90.15}&74.78&81.38&61.93&51.66&56.08&-&-&-\\
  hFT-Transformer&&5.5M&83.36&\underline{82.00}&\underline{82.67}&86.63&\underline{83.75}&\underline{85.07}&\underline{67.18}&\underline{65.06}&\underline{66.03}&\underline{48.75}&\underline{47.21}&\underline{47.92}\\
  hFT-Transformer&\checkmark&5.5M&83.68&\textbf{82.11}&\textbf{82.89}&\underline{86.72}&\textbf{83.81}&\textbf{85.14}&\textbf{67.51}&\textbf{65.36}&\textbf{66.34}&\textbf{49.05}&\textbf{47.48}&\textbf{48.20}\\
  \hline
 \end{tabular}
 \caption{Evaluation results on MAPS test dataset (P: precision, R: recall, \textbf{bold}: best score, \underline{underline}: second best score)}
 \label{tab:table_1}%Table 1
 \end{center}
\end{table*}
\begin{table*}
 \begin{center}
 \small
 \tabcolsep=3pt
 \begin{tabular}{ccc|cccccccccccc}
  \hline
  \multirow{2}{*}{Method}&half&\multirow{2}{*}{Params}&\multicolumn{3}{c}{Frame}&\multicolumn{3}{c}{Note}&\multicolumn{3}{c}{Note w/ Offset}&\multicolumn{3}{c}{Note w/ Offset\&Velocity}\\
  &stride&&P(\%)&R(\%)&F1(\%)&P(\%)&R(\%)&F1(\%)&P(\%)&R(\%)&F1(\%)&P(\%)&R(\%)&F1(\%)\\
  \hline
  Seq2Seq\cite{HAWTHORNE2021:01}&&54M&-&-&-&-&-&96.01&-&-&83.94&-&-&82.75\\
  HPT-T\cite{OU2022:01}&&-&-&-&90.09&97.88&\textbf{96.72}&96.77&84.13&82.31&83.20&82.85&81.07&81.90\\
  Semi-CRFs\cite{YAN2021:01}&&9M&\textbf{93.79}&88.36&90.75&98.69&93.96&96.11&90.79&86.46&88.42&89.78&85.51&87.44\\
  HPPNet-sp\cite{WEI2022:01}&&1.2M&92.79&\underline{93.59}&\underline{93.15}&98.45&\underline{95.95}&97.18&84.88&82.76&83.80&83.29&81.24&82.24\\
  hFT-Transformer&&5.5M&92.62&93.43&93.02&\underline{99.62}&95.41&\underline{97.43}&\underline{92.32}&\underline{88.48}&\underline{90.32}&\underline{91.21}&\underline{87.44}&\underline{89.25}\\
  hFT-Transformer&\checkmark&5.5M&\underline{92.82}&\textbf{93.66}&\textbf{93.24}&\textbf{99.64}&95.44&\textbf{97.44}&\textbf{92.52}&\textbf{88.69}&\textbf{90.53}&\textbf{91.43}&\textbf{87.67}&\textbf{89.48}\\
  \hline
 \end{tabular}
 \caption{Evaluation results on MAESTRO v3.0.0 test dataset}
 \label{tab:table_2}%Table 2
 \end{center}
\end{table*}

\subsection{Inference}\label{subsec:inference}
At inference time, we use \textit{output\_2nd} as the final output.
%The threshold for \textit{frame} is set to 0.5.
We set the threshold for \textit{frame} as 0.5.
For note-wise events (\textit{onset}, \textit{offset}, and \textit{velocity}), the outputs in each pitch-frame grid are converted to a set containing note-wise onset, offset, and velocity following Kong et al.'s \textit{Algorithm 1}~\cite{KONG2021:01} in five steps shown below:

\begin{description}[style=unboxed,leftmargin=0cm,topsep=0.1em,partopsep=0cm,itemsep=0.3em,parsep=0cm]
\item[Step 1. onset detection:] find a local maximum in \textit{onset} with a value at least 0.5. Then calculate the precise onset time using the values of the adjacent three frames~\cite{KONG2021:01}.
\item[Step 2. velocity:] If an onset is detected in Step 1, extract the \textit{velocity} value at the frame. If the value is zero, then discard both onset and velocity at this frame.
\item[Step 3. offset detection with \textit{offset}:] find a local maximum in \textit{offset} with a value at least 0.5. Then calculate the precise offset time using the values of the adjacent three frames~\cite{KONG2021:01}.
\item[Step 4. offset detection with \textit{frame}:] choose the frame that is nearest to the detected onset which has a \textit{frame} value below 0.5.
\item[Step 5. offset decision:] choose the smaller value between the results of Step 3 and 4.
\end{description}

\begin{figure}[t]
 \centering
 \includegraphics[width=0.99\columnwidth]{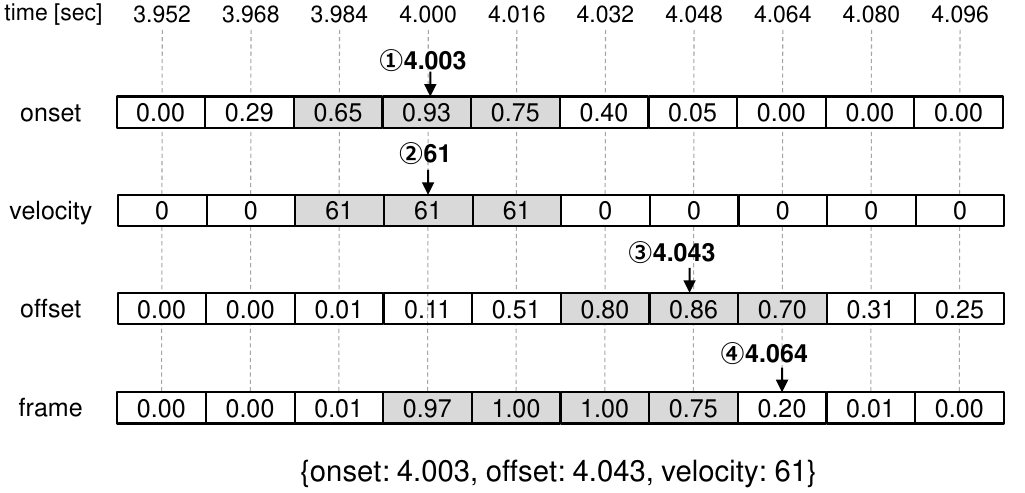}
 \caption{Example of conversion from grid-wise values to note-wise values}
 \label{fig:figure_5}%Figure 5
\end{figure}

An example is shown in Figure \ref{fig:figure_5}.
The \textit{onset} is 4.003, and the \textit{velocity} is 61.
For \textit{offset}, the direct estimation from \textit{offset} is 4.043, and that estimated via \textit{frame} is 4.064.
Thus, we choose 4.043 as \textit{offset}.
Finally, we obtain a note with \{\textit{onset}: 4.003, \textit{offset}: 4.043, \textit{velocity}: 61\} in the output.

%
% 4-4. Metrics
%
\subsection{Metrics}\label{subsec:metrics}
We evaluate the performance of our proposed method with frame-level metrics (\textit{Frame}) and note-level metrics (\textit{Note}, \textit{Note with Offset}, and \textit{Note with Offset \& Velocity}) with the standard precision, recall, and F1 scores.
We calculated these scores using \texttt{mir\_eval} library~\cite{RAFFEL2014:01} with its default settings.
The scores were calculated per recording, and the mean of these per-recording scores was presented as the final metric for a given collection of pieces, as explained in Hawthorne et al.~\cite{HAWTHORNE2018:01}.

%
% 4-5. Results
%
\subsection{Results}\label{subsec:results}
Tables \ref{tab:table_1} and \ref{tab:table_2} show the scores on the test sets of MAPS and MAESTRO datasets.
The numbers of parameters in these Tables are referred from \cite{WEI2022:01,KELZ2019:01}.
For the MAPS dataset, our proposed method outperformed the other methods in F1 score for all metrics.
For the MAESTRO dataset, our proposed method outperformed the other methods in F1 score for \textit{Note}, \textit{Note with Offset}, and \textit{Note with Offset \& Velocity}.
Furthermore, our method with the half-stride strategy which is mentioned in  \ref{subsec:inference_stride} outperformed other methods in all metrics.
In contrast, the two state-of-the-art methods for MAESTRO, which are Semi-CRFs \cite{YAN2021:01} and HPPNet-sp \cite{WEI2022:01}, performed well only on a subset of the metrics.

The results suggest that the proposed two-level hierarchical frequency-time Transformer structure is promising for AMT.

%
% 4-6. Ablation Study
%
\begin{table*}
\begin{center}
 \small
 \tabcolsep=3pt
 \begin{tabular}{c|ccccc}
  \hline
  \multirow{2}{*}{Model}&\multicolumn{3}{c}{1st-Hierarchy}&2nd-Hierarchy&\multirow{2}{*}{Output}\\
  &Convolutional block&1st Transformer encoder&Converter&2nd Transformer encoder&\\
  \hline
  1-F-D-T{\dag}&1-D (time axis)&Frequency axis&Transformer Decoder&Time axis&output\_2nd\\
  1-F-D-N&1-D (time axis)&Frequency axis&Transformer Decoder&n/a&output\_1st\\
  2-F-D-T&2-D&Frequency axis&Transformer Decoder&Time axis&output\_2nd\\
  1-F-L-T&1-D (time axis)&Frequency axis&Linear&Time axis&output\_2nd\\
 \end{tabular}
 \caption{Model variations for ablation study (\dag: the proposed method, hFT-Transformer)}
 \label{tab:table_3}%Table 3
 \end{center}
\end{table*}
\begin{table*}
 \begin{center}
 \small
 \tabcolsep=3pt
 \begin{tabular}{cc|cccccccccccc}
  \hline
  \multirow{2}{*}{Model}&\multirow{2}{*}{Params}&\multicolumn{3}{c}{Frame}&\multicolumn{3}{c}{Note}&\multicolumn{3}{c}{Note w/ Offset}&\multicolumn{3}{c}{Note w/ Offset\&Velocity}\\
  &&P(\%)&R(\%)&F1(\%)&P(\%)&R(\%)&F1(\%)&P(\%)&R(\%)&F1(\%)&P(\%)&R(\%)&F1(\%)\\
  \hline
  1-F-D-T{\dag}&5.5M&\underline{93.61}&\textbf{88.71}&\textbf{91.09}&98.81&\textbf{94.81}&\textbf{96.72}&\textbf{86.18}&\textbf{82.81}&\textbf{84.42}&\textbf{77.47}&\textbf{74.55}&\textbf{75.95}\\
  1-F-D-N&3.9M&92.85&87.49&90.09&\underline{99.01}&\underline{93.24}&\underline{95.95}&82.67&78.06&80.23&\underline{73.89}&\underline{69.90}&\underline{71.78}\\
  2-F-D-T&6.1M&75.49&61.08&67.52&97.03&19.68&31.10&64.07&13.28&20.88&42.11&8.57&13.50\\
  1-F-L-T&3.4M&\textbf{93.71}&\underline{88.42}&\underline{90.99}&\textbf{99.11}&92.90&95.79&\underline{85.77}&\underline{80.56}&\underline{82.98}&71.66&67.32&69.34\\
  \hline
 \end{tabular}
 \caption{Evaluation results of ablation study on MAPS validation dataset}
 \label{tab:table_4}%Table 4
 \end{center}
\end{table*}
\subsection{Ablation Study}\label{subsec:ablation_study}
To investigate the effectiveness of each module in our proposed method, we trained various combinations of those modules using the MAPS training set and evaluated them using the MAPS validation set.
The variations are shown in Table \ref{tab:table_3}.
In this study, we call our proposed method \textit{1-F-D-T}, which means it consists of the \textit{1}-D convolution block, the first Transformer encoder in the \textit{F}requency axis, the Transformer \textit{D}ecoder, and the second Transformer encoder in the \textit{T}ime axis.
Table \ref{tab:table_4} shows evaluation results for each variation.

\textbf{Second Transformer encoder in time axis.}
%1-F-D-T vs 1-F-D-N
To verify the effectiveness of the second Transformer encoder, we compared the 1-F-D-T and the model without the second Transformer encoder (1-F-D-N).
For the 1-F-D-N model, we use \textit{output\_1st} in both training and inference stages as the final output.
The result indicates that the second Transformer encoder improved \textit{Note with Offset} performance, in which the F1 score is 84.42 for 1-F-D-T and 80.23 for 1-F-D-N.
This shows the effectiveness of the second Transformer encoder as it provides an extra pass to model the temporal dependency of acoustic features, which is presumably helpful in offset estimation.

\textbf{Compelxity of convolutional block.}
%1-F-D-T vs 2-F-D-T
To investigate how the complexity of the convolutional block affects the AMT performance, we compared the 1-F-D-T model and the model that replaces the 1-D convolutional block with a 2-D convolutional block (2-F-D-T).
Surprisingly, the result shows that the performance of the 2-F-D-T model is significantly worse than that of the 1-F-D-T model.
This is probably because the two modules working on the spectral dependency do not cohere with each other.
The 2-D convolutional block may over aggregate the spectral information thus resulting into an effectively lower frequency resolution. Then, the Transformer encoder can only evaluate the spectral dependency over an over-simplified feature space, causing the performance degradation. 

\textbf{Converter.}
%1-F-D-T vs 1-F-L-T
We used a Transformer decoder to convert the dimension in the frequency axis from $F$ to $P$.
In contrast, almost all of the existing methods used a linear module to achieve this.
We compared the performance of the 1-F-D-T model to a model with the Transfomer decoder replaced with a linear converter (1-F-L-T).
The result indicates that the 1-F-D-T model outperformed the 1-F-L-T model in F1 score for all four metrics.
Especially, the difference in \textit{Note with Offset and Velocity} is large (75.95 for the 1-F-D-T model and 69.34 for the 1-F-L-T model in F1 score).
This suggests that using a Transformer decoder as converter is an effective way of improving the performance, although the side effect is the increase of model size. 

\begin{figure}
 \centering
 \includegraphics[width=0.99\columnwidth]{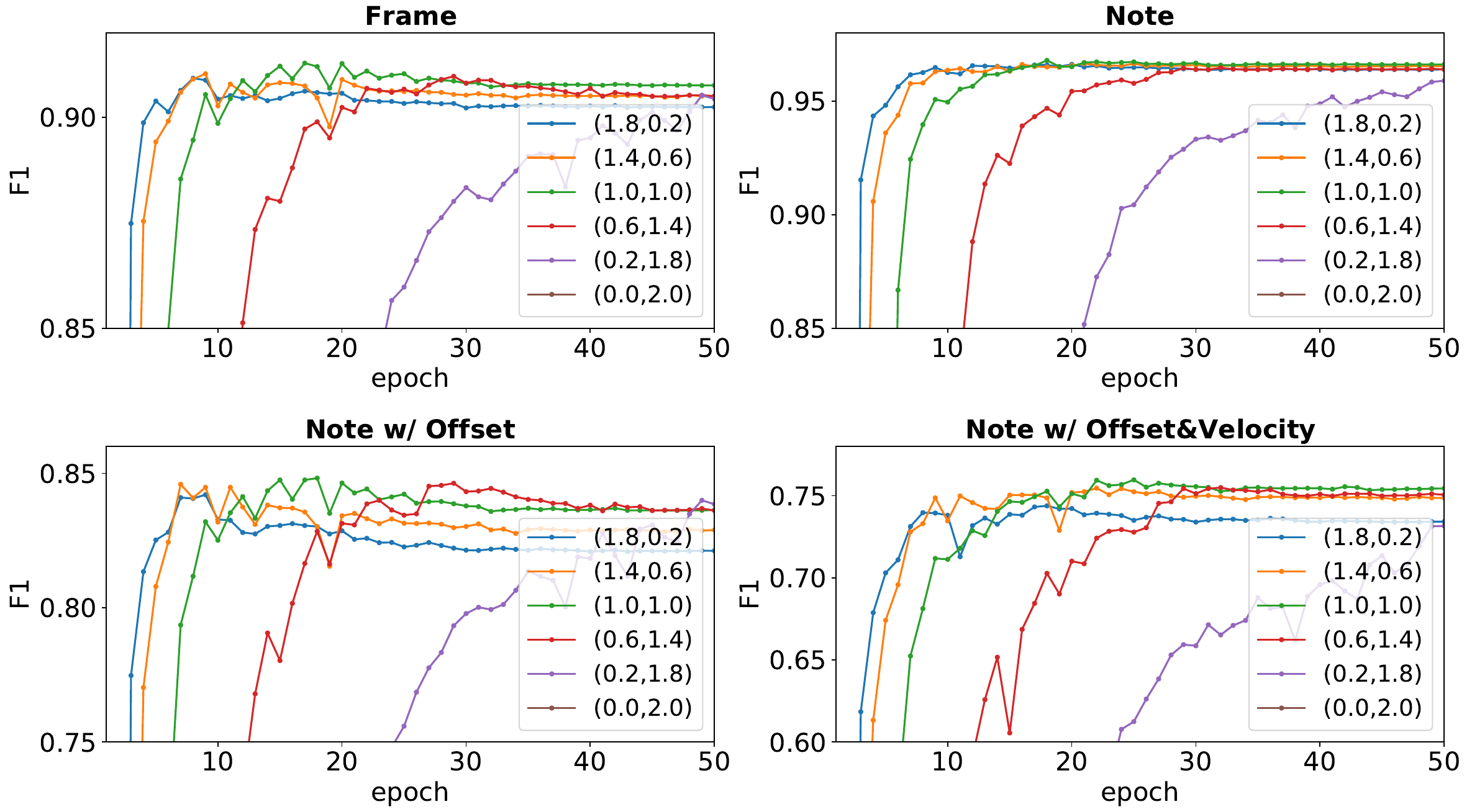}
 \caption{Performance of 1-F-D-T model trained with six-pairs of coefficients of loss functions}
 \label{fig:figure_6}% Figure 6
\end{figure}

We also investigated how the coefficients for the loss functions, $\alpha_{\mathrm{1st}}$ and $\alpha_{\mathrm{2nd}}$ in Eqn (\ref{eqn:loss_combination}), affect the performance.
We investigated six pairs of coefficients of loss functions ($\alpha_{\mathrm{1st}}$, $\alpha_{\mathrm{2nd}}$) in Eqn (\ref{eqn:loss_combination}), i.e., (1.8, 0.2), (1.4, 0.6), (1.0, 1.0), (0.6, 1.4), (0.2, 1.8), and (0.0, 2.0), for the 1-F-D-T model.
Figure \ref{fig:figure_6} shows the F1 scores of \textit{frame}, \textit{onset}, \textit{offset}, and \textit{velocity} evaluated on the MAPS validation set in each epoch.
These results indicate that the (1.0, 1.0) pair yields the best score.
It also shows that the training converges faster when $\alpha_{\mathrm{1st}}$ is larger than $\alpha_{\mathrm{2nd}}$.
Importantly, if we omit the \textit{output\_1st}, which is the case when training with the pair (0.0, 2.0), the training loss did not decrease much.
Therefore, the F1 score stays around 0\% and thus cannot be seen in Figure \ref{fig:figure_6}.
This suggests that it is crucial to use both losses, \textit{output\_1st} and \textit{output\_2nd} in our proposed method.

%
% 5. Conclusion
%
\section{Conclusion}\label{sec:conclusion}
In this work, we proposed \textit{hFT-Transformer}, an automatic piano transcription method that uses a two-level hierarchical frequency-time Transformer architecture.
The first hierarchy consists of a 1-D convolutional block in the time axis, a Transformer encoder and a Transformer decoder in the frequency axis, and the second hierarchy consists of a Transformer encoder in the time axis.
The experiment result based on two well-known piano datasets, MAPS and MAESTRO, revealed that our two-level hierarchical architecture works effectively and outperformed other state-of-the-art methods in F1 score for frame-level and note-level transcription metrics.
For future work, we would like to extend our method to other instruments and multi-instrument settings.

%
% 6. Acknowledgments
%
\section{Acknowledgments}\label{sec:acknowledgments}
We would like to thank Giorgio Fabbro and Stefan Uhlich for their valuable comments while preparing this manuscript.
We are grateful to Kin Wai Cheuk for his dedicated support in preparing our github repository.

%
% Reference
%
% For bibtex users:
\bibliography{AMT_reference}

% For non bibtex users:
%\begin{thebibliography}{citations}
% \bibitem{Author:17}
% E.~Author and B.~Authour, ``The title of the conference paper,'' in {\em Proc.
% of the Int. Society for Music Information Retrieval Conf.}, (Suzhou, China),
% pp.~111--117, 2017.
%
% \bibitem{Someone:10}
% A.~Someone, B.~Someone, and C.~Someone, ``The title of the journal paper,''
%  {\em Journal of New Music Research}, vol.~A, pp.~111--222, September 2010.
%
% \bibitem{Person:20}
% O.~Person, {\em Title of the Book}.
% \newblock Montr\'{e}al, Canada: McGill-Queen's University Press, 2021.
%
% \bibitem{Person:09}
% F.~Person and S.~Person, ``Title of a chapter this book,'' in {\em A Book
% Containing Delightful Chapters} (A.~G. Editor, ed.), pp.~58--102, Tokyo,
% Japan: The Publisher, 2009.
%
%
%\end{thebibliography}

\end{document}